\newif\ifsingle

\newif\ifproofs

\ifsingle
\documentclass[11pt,draftclsnofoot, onecolumn]{IEEEtran}		
\else		
\documentclass[conference]{IEEEtran} 
\fi

\usepackage{times}
\usepackage{amsmath,dsfont}
\usepackage{amssymb,amsthm}
\usepackage{epsfig,verbatim} 
\usepackage{setspace}
\usepackage{color}
\usepackage{xcolor}
\usepackage{cite}
\usepackage{epstopdf}
\usepackage{graphics}
\usepackage{graphicx}
\usepackage{accents}
\usepackage{acronym}
\usepackage[bookmarks,colorlinks]{hyperref}
\usepackage{booktabs}
\usepackage{mathtools}
\usepackage{float}
\usepackage[linesnumbered,ruled,vlined]{algorithm2e}
\usepackage{algpseudocode}
\usepackage{enumitem}
\usepackage{bm}
\usepackage[section]{placeins}
\usepackage{tablefootnote}
\usepackage{threeparttable}
\usepackage{subcaption}

\newcommand{\removelatexerror}{\let\@latex@error\@gobble}
\newcommand{\myVec}[1]{{\boldsymbol{#1}}}
\newcommand{\myMat}[1]{{\boldsymbol{#1}}}
\newcommand{\mySet}[1]{\mathcal{#1}}

\acrodef{fc}[FC]{fully-connected}
\acrodef{bc}[BC]{broadcast channel}
\acrodef{mac}[MAC]{multiple access channel}  
\acrodef{adc}[ADC]{analog-to-digital converter}  
\acrodef{csi}[CSI]{channel state information} 
\acrodef{snr}[SNR]{signal-to-noise ratio}
\acrodef{sinr}[SINR]{signal-to-interference-and-noise ratio}
\acrodef{tx}[TX]{Transmitter}  
\acrodef{mimo}[MIMO]{multiple-input multiple-output}
\acrodef{mse}[MSE]{mean-squared error}
\acrodef{lmmse}[LMMSE]{linear minimum mean-squared error}
\acrodef{pdf}[PDF]{probability density function}
\acrodef{rv}[RV]{random variable} 
\acrodef{isi}[ISI]{intersymbol interference}  
\acrodef{awgn}[AWGN]{additive white Gaussian noise} 
\acrodef{lti}[LTI]{linear time-invariant}  
\acrodef{ut}[UT]{user terminal} 
\acrodef{mmw}[mmWave]{millimeter wave}
\acrodef{dma}[DMA]{dynamic metasurface antenna}
\acrodef{ofdm}[OFDM]{orthogonal frequency division multiplexing}
\acrodef{ofdma}[OFDMA]{orthogonal frequency division multiple access}
\acrodef{dnn}[DNN]{deep neural network}
\acrodef{gnn}[GNN]{graph neural network}
\acrodef{ml}[ML]{machine learning}
\acrodef{dl}[DL]{deep learning}
\acrodef{mlp}[MLP]{Multilayer Perceptron}
\acrodef{sgd}[SGD]{stochastic gradient descent}
\acrodef{bpsk}[BPSK]{binary phase shift keying}
\acrodef{pgd}[PGD]{projected gradient descent}
\acrodef{ber}[BER]{bit error rate}
\acrodef{csi}[CSI]{channel state information}
\acrodef{map}[MAP]{maximum a-posteriori probability}
\acrodef{ct}[CT]{continuous-time}
\acrodef{manet}[MANET]{mobile ad hoc network}
\acrodef{noma}[NOMA]{non-orthogonal multiple access}
\acrodef{sic}[SIC]{successive interference cancellation}
\acrodef{mmse}[MMSE]{minimum mean squared error}
\acrodef{siso}[SISO]{single input single output}
\acrodef{gat}[GAT] {Graph Attention Network}
\acrodef{adam}[ADAM]{Adaptive Moment Estimation}
\acrodef{film}[FiLM]{Feature-wise Linear Modulation}

\definecolor{blue}{rgb}{0,0,1}

\DeclareMathOperator*{\argmax}{argmax}

\ifsingle

\else

\fi 

\usepackage[all=normal,paragraphs=tight,floats=normal,mathspacing=normal,wordspacing=tight,charwidths=tight,mathdisplays=normal,leading=normal]{savetrees}

\IEEEoverridecommandlockouts

\title{Decentralized Multi-Channel MANET Power Optimization Using Graph Neural Networks
}
\author{
	\IEEEauthorblockN{Tomer Alter, Nir Shlezinger, and Michael Segal\\
    \thanks{
The authors are with the ECE School, Ben-Gurion University of the Negev, Israel (e-mail: tomeralt@post.bgu.ac.il; \{nirshl; segal\}@bgu.ac.il).  The work of this paper has been partially funded by  Israeli Science Foundation (Grant No. 465/22),
US Army Research Office under Grant Number W911NF-22-1-0225,  and the European Research Council (ERC) under the ERC starting grant nr. 101163973 (FLAIR). 
}
	}

}
\allowdisplaybreaks
\begin{document}

\maketitle 
\begin{abstract} 
The increasing demand for \acp{manet} calls for decentralized mechanisms that can allocate transmit power across nodes and channels under stringent resource constraints. Existing optimization-based approaches, however,   do not account for expected settings where each link includes multiple channels (e.g., multi-band signaling). Motivated by recent advances in machine learning for distributed optimization, we propose \emph{MANET-GNN}, a \ac{gnn}-based algorithm for decentralized power allocation in multi-channel \acp{manet}. MANET-GNN explicitly exploits the network topology, scales efficiently with the number of nodes and frequency bands, generalizes across topologies and channel conditions, and enables near-instantaneous inference suitable for real-time deployment. Our design builds on a constrained optimization formulation and employs a dedicated \ac{gnn} architecture inspired by message passing, trained via an unsupervised procedure that is robust to noisy channel state information. Numerical evaluations  demonstrate that MANET-GNN achieves high-throughput multi-channel communication across diverse \ac{manet} scenarios. 
\end{abstract}


\acresetall
\section{Introduction}

\Ac{manet} is a popular framework for studying and developing {flexible  decentralized communication systems} without fixed infrastructure~\cite{tavli2006mobile}. Their ability to self-organize and adapt makes them attractive for a wide range of applications, including vehicular communications, industrial IoT, emergency response networks, and tactical operations. In these scenarios, devices must dynamically form multi-hop topologies and collaboratively manage shared wireless resources under strict latency and energy constraints~\cite{kafetzis2022software}.

In various \ac{manet} technologies, communication is enabled via \emph{multiple channels} for each link, either due to heterogeneous communication technologies, spectrum partitioning, or the availability of orthogonal frequency bands~\cite{xie2021multi,karabulut2022novel,chen2024joint}. While leveraging multiple channels can significantly improve throughput and reliability, it also introduces new challenges. Since \ac{manet} devices are often power-limited and operate in rapidly changing environments, efficient allocation of transmit power across both \emph{links} and \emph{channels} becomes critical for maintaining high spectral efficiency and reliable connectivity. However, the successful exploitation of multi-channel diversity requires dedicated mechanisms for distributed power allocation that can scale to large networks.

A wide variety of optimization-based solutions have been proposed to address the problem of power allocation in \acp{manet}~\cite{kanellopoulos2020survey}. These methods span multiple categories, including adaptations of radio state operational modes~\cite{singh1998pamas, ye2002smac}, adaptive load balancing~\cite{kim2002power}, location- and multicast-based routing~\cite{karp2000gpsr,li2001learn}, and cross-layer optimization frameworks~\cite{mafirabadza2016efficient}.  Despite their diversity, most of these works have been developed under the assumption of single-channel \acp{manet}, where each link operates over a single communication resource. Consequently, they do not naturally extend to the multi-channel setting, where each link may exploit multiple subcarriers or physical channels. This limitation highlights the need for  decentralized optimization frameworks that can effectively scale to the more demanding multi-channel \ac{manet} scenario.



The growing success of deep learning has led to an alternative paradigm for tackling challenging distributed processing tasks using data-driven tools. A popular framework is distributed reinforcement learning, in which each agent is equipped with a local \ac{dnn} trained to autonomously adjust its decisions for decentralized resource allocation~\cite{randall2022grows, liu2024graph}. However, such approaches typically treat nodes independently and remain agnostic to the underlying graph structures, thereby limiting their ability to generalize across unseen network topologies. An alternative direction is the \emph{learn-to-optimize} paradigm~\cite{shlezinger2022model}, which casts tasks such as resource allocation as an optimization problem and trains \ac{dnn} or unfolded optimizers to rapidly generate valid solutions. This paradigm has shown success in centralized resource allocation problems~\cite{alter2025rapid} as well as in decentralized consensus optimization~\cite{noah2024distributed}. In the context of decentralized processing of graph signals, \acp{gnn} have been demonstrated to provide graph-aware architectures that can be trained for distributed optimization, with architectures designed for  power allocation in single-channel settings~\cite{eisen2020optimal}, over-the-air aggregation~\cite{gu2023graph}, link sparsification~\cite{zhao2025distributed}, and wireless federated learning enhancement~\cite{li2022power}. These advances motivate the adaptation of \ac{gnn}-based optimization methodologies to the challenging scenario of distributed resource allocation in multi-channel \acp{manet}.

In this work, we develop a GNN-based algorithm, termed \emph{MANET-GNN}, for decentralized power allocation in multi-channel \acp{manet}. Our design $(i)$ explicitly exploits the \emph{graph topology} of the network to capture the structured interactions among nodes; $(ii)$ scales efficiently with both the number of devices and the available frequency bands; $(iii)$ generalizes across different network topologies and channel conditions, and $(iv)$ enables \emph{near-instantaneous inference}, making it suitable for real-time deployment in dynamic MANET environments.

Inspired by learned optimization techniques~\cite{chen2022learning}, our approach begins by formulating the multi-channel resource allocation problem as a constrained optimization task, aiming to maximize throughput between a given transmitter–receiver pair in the \ac{manet}. While this optimization problem is inherently non-convex and requires a centralized solver, we leverage it to guide the training of our dedicated MANET-GNN architecture that requires solely local \ac{csi}, while being inherently amenable to distributed execution over varying graphs and channel conditions. Our unsupervised training procedure facilitates learning based on the optimization objective by incorporating a dedicated smooth-min approximation.  We then conduct  numerical evaluations,  demonstrating that MANET-GNN consistently enables high-throughput multi-channel communication across diverse \ac{manet} topologies, achieving performance that is competitive with fully centralized optimization while operating in a fully decentralized manner.


The remainder of this paper is organized as follows: Section~\ref{sec:System} introduces the system model and problem formulation; Section~\ref{sec:Decentralized} presents the proposed decentralized learned optimization framework, which we evaluate in Section~\ref{sec:experiments}.
Finally, Section~\ref{sec:Conclusions} concludes the paper.

\section{System Model}
\label{sec:System}
In this section, we formulate the system model for multi-channel \acp{manet}. We commence with presenting the communication system in Subsection~\ref{sec:Com system}, and 
formulate the decentralized power allocation optimization problem in Subsection~\ref{sec:Problem}, which constitutes the starting point  for MANET-GNN.

\subsection{Communication System Model}
\label{sec:Com system}

\begin{figure}[t!]
  \includegraphics[width=0.95\columnwidth,keepaspectratio]{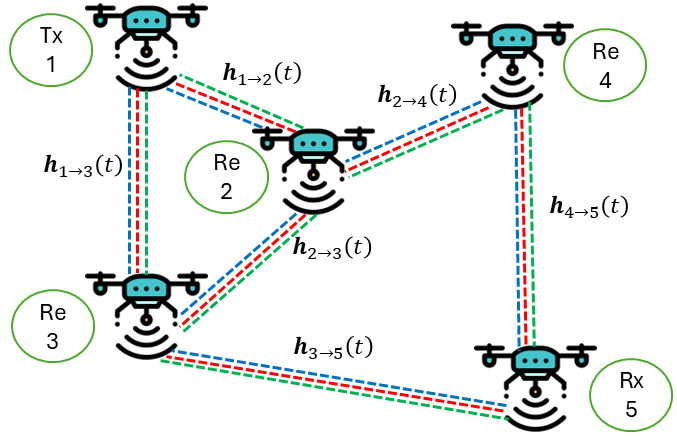}
  \caption{Multi-channel \ac{manet}, $B=3$, $|\mySet{V}|=5$. }
  \label{fig:example grap}
\end{figure}

We consider a block-fading multi-hop \ac{manet} with reciprocal links comprised of $B$ channels (e.g., multiple physical channels, or multi-carrier signaling).  We model the network topology as an undirected graph $\mathcal{G} = (\mathcal{V}, \mathcal{E})$, where $\mathcal{V}$ represents the set of nodes and $\mathcal{E}$ represents the set of wireless links. In each block of index $t$, a link $(i,j)\in\mathcal{E}$ is represented using a $B\times 1$ vector $[h_{i \rightarrow j}^{(1)}(t), \ldots, h_{i \rightarrow j}^{(B)}(t)]$, with $h_{i \rightarrow j}^{(b)}(t)\in \mathbb{C}$ being the realization of the $b$th channel in the link.  In each block $t$, one node $u^{\rm Tx}\in \mySet{V}$ acts as the \emph{source} and another $u^{\rm Rx}\in \mySet{V}$ as the \emph{destination}, while the remaining nodes serve as relays (see Fig.~\ref{fig:example grap}).  Both the wireless channels and the network topology vary over time in a blockwise manner, capturing the dynamic nature of ad hoc networks.

Let $y_{i \rightarrow j}^{(b)}(t)$ denote the received signal over link $(i,j)\in \mySet{E}$ on channel $b$ during block $t$. The received signal is given by
\begin{align} 
    y_{i \rightarrow j}^{(b)}(t) 
    &=  h_{i \rightarrow j}^{(b)}(t)\, p_{i \rightarrow j}^{(b)}(t)\, s_i^{(b)}(t) 
    + w_{i \rightarrow j}^{(b)}(t),
    \label{eq:received message b}
\end{align}
where $s_i^{(b)}(t)$ is the unit amplitude signal transmitted by the $i$th node on channel $b$, while $w_{i \rightarrow j}^{(b)}(t)\sim \mySet{CN}(0,\sigma_b^2)$ is \ac{awgn}. 
For convenience, the key variables used in the system model are summarized in Table~\ref{tab:variables}.

\begin{table}
    \centering
    \caption{Key variables and parameters}
    \label{tab:variables}
    \resizebox{\columnwidth}{!}{%
    \begin{tabular}{ll}
        \hline
        \textbf{Symbol} & \textbf{Definition} \\
        \hline
        $w_{{i \rightarrow j}}^{(b)}(t)$ & AWGN noise at link $(i,j)$ on channel $b$ \\
        $h_{i \rightarrow j}^{(b)}(t)$ & Channel coefficient between nodes $i$ and $j$ on channel $b$ \\
        $p_{i \rightarrow j}^{(b)}(t)$ & Power allocated by node $i$ to node $j$ on channel $b$ \\
        $s_i^{(b)}(t)$ & Transmitted signal from node $i$ on channel $b$ \\
        $\mySet{N}(j)$ & Set of neighboring nodes of node $j$ \\
        \hline
    \end{tabular}%
    }
\end{table}

\subsection{Power Allocation}
\label{sec:Problem}

We aim to determine the power allocation across $B$ frequency bands for all links 
in the \ac{manet} (i.e., $\{ p_{i \rightarrow j}^{(b)}(t)\}$ for all $(i,j)\in \mySet{E}$), such that the end-to-end communication rate between the source and destination is maximized 
while satisfying per-node power constraints. We next cast this task  as an optimization setting, based on which we formulate our power allocation problem.

\subsubsection{Constrained Optimization}
Since we operate on a signal block, we next omit the block index $t$ for brevity, and stack the power allocation as the $B\times |\mySet{V}| \times |\mySet{V}|$ tensor $\myMat{P}$, with $[\myMat{P}]_{b,i,j} =  p_{i \rightarrow j}^{(b)}$. 
For a given frequency band $b$ and link $(i\!\to\!j)\in\mySet{E}$, the achievable rate for the \ac{awgn} channel in \eqref{eq:received message b} is  given by:
\begin{align}
    R_{i\to j}^{(b)} 
    = \log_2 \!\left( 
        1 + \frac{\big|h_{i\to j}^{(b)}\big|^2 \big(p_{i\to j}^{(b)}\big)^2}{\sigma_b^2} 
    \right).
\end{align} 
Let $\Phi_b$ denote the set of all available paths between the source and destination on band $b$.  
A path $\phi \in \Phi_b$ is defined as an ordered sequence of edges connecting the source to the destination 
in the graph $\mathcal{G}$, where an edge $(i\!\to\!j)$ exists only if $\big|h_{i\to j}^{(b)}\big| > 0$.  
The achievable rate of a path $\phi$ on band $b$ is determined by its \emph{bottleneck link}, i.e., 
\(\displaystyle R_{\phi}^{(b)} = \min_{(i\to j)\in\phi} R_{i\to j}^{(b)}\).

We impose two constraints on the transmit powers:
\begin{itemize}
    \item \textbf{Per-node power limit:}  the overall power allocated by each node $i\in \mySet{V}$ holds
    \( 
        { \sum_{b=1}^B \sum_{(j) \in \mySet{N}(i)} 
        \big( p_{i\to j}^{(b)} \big)^2 } \leq 1
    \).
    \item \textbf{Non-negativity:}  all settings hold
    \(\displaystyle p_{i\to j}^{(b)} \geq 0\).
\end{itemize}
Accordingly, we formulate the task of power allocation for optimizing end-to-end communications in multi-channel \acp{manet} as the following optimization problem 
\begin{align} 
    \myMat{P}^\star 
    = \argmax_{\myMat{P}\in\mySet{P}}  
      \sum_{b=1}^{B} 
      \max_{\phi \in \Phi_b} \, 
      \min_{(i\to j) \in \phi} R_{i\to j}^{(b)},
    \label{eqn: best power allocation}
\end{align}
where the feasible set of power allocations is defined as:
\begin{equation}
\label{eqn:PSet}
    \mySet{P} = \Big\{ 
    \myMat{P} \in [0,1]^{B \times n \times n} : 
    \big\|[\myMat{P}]_{:,i,:}\big\|_2 \leq 1,\ \forall i\in\mySet{V} 
    \Big\}.
\end{equation}
The formulation in \eqref{eqn: best power allocation} allows the usage of different paths in different channels (as the path with the highest rate $\phi \in \Phi_b$ is selected in each channel $b$).

\subsubsection{Problem Formulation}
Our goal is to design a policy for setting the power allocations $\myMat{P}$ based on the formulated optimization objective in \eqref{eqn: best power allocation}. While the power allocation problem in \eqref{eqn: best power allocation} is formulated as a centralized optimization setting, i.e., a mapping of the full \ac{manet} \ac{csi} $\{h_{i \to j}^{(b)}\}_{(i,j)\in \mySet{E}, b \in \{1,\ldots B\}}$ into $\myMat{P}^{\star}$, we aim to design a method that meets the following requirements: 
\begin{enumerate}[label={R\arabic*}]
    \item \label{itm:Decent} {\em Decentralized operation}, i.e., each node $i$ sets its own $\{p_{i \to j}^{(b)}\}$ based on its local \ac{csi} $\{\hat{h}_{i \to j}^{(b)}\}_{(j)\in \mySet{N}(i), b \in \{1,\ldots B\}}$. 
    \item  \label{itm:Messages} {\em Limited latency optimization}, where each node $i$ is allowed to exchange at most $L$ messages with its neighbors $\mySet{N}(i)$. 
    \item \label{itm:transfer} The method should be applicable on {\em different  topologies}. 
      \item \label{itm:Noisy} The local \ac{csi}  $\{\hat{h}_{i \to j}^{(b)}\}_{(j)\in \mySet{N}(i), b \in \{1,\ldots B\}}$, may be a {\em noisy estimate} of the actual \ac{csi}.  
\end{enumerate}
To cope with \ref{itm:Decent}-\ref{itm:Noisy}, we assume access to \ac{csi} from various \acp{manet} during design, represented by the data set
\begin{equation}
    \mySet{D} = \left\{\{h_{i \to j, d}^{(b)}\}_{(i,j)\in \mySet{E}_d, b \in \{1,\ldots B\}}, \mySet{G}_d = (\mySet{V}_d, \mySet{E}_d)\right\}_{d=1}^{|\mySet{D}|}.
    \label{eqn:DataSet}
\end{equation}
Note that \eqref{eqn:DataSet} does not contain 'ground-truth'  allocations, and that its \acp{csi} come from different \acp{manet} with different topologies. 


\section{Decentralized Learned Optimization}
\label{sec:Decentralized}

In this section, we present our proposed decentralized learned optimization framework. 
To address the challenging requirements \ref{itm:Decent}--\ref{itm:Noisy}, we propose to tune the power allocation policy by leveraging the optimization formulation in \eqref{eqn: best power allocation} through \emph{learned optimization} tools. Our design builds upon the recent empirical success of \acp{gnn} in solving learned optimization tasks~\cite{shen2020graph, randall2022grows}, exploiting their inherent ability to operate in a decentralized manner (\ref{itm:Decent}) while naturally adapting to different network topologies (\ref{itm:transfer})~\cite{corso2024graph}. Specifically, we introduce a dedicated \ac{gnn} architecture, termed \emph{MANET-GNN}, that is inspired by message-passing networks~\cite{feng2022powerful} and explicitly constrains the number of message exchanges to meet the latency requirement in \ref{itm:Messages} (see Subsection~\ref{sec:arch}). Furthermore, to enhance robustness under imperfect channel knowledge as noted in \ref{itm:Noisy}, we devise in Subsection~\ref{sec:training} an unsupervised training procedure that directly optimizes the power allocation objective as a discriminative machine learning model~\cite{shlezinger2022discriminative} while accounting for the expected noisiness of the local \ac{csi}. This design yields a principled framework that not only aligns with the optimization formulation but also satisfies the practical requirements of decentralized and real-time operation in multi-channel \acp{manet}.

\subsection{MANET-GNN}
\label{sec:arch}
\subsubsection{Gated GNN}
Our main building block is a learned message passing module, which updates each node's local embedding based on a single round of information exchange with its neighbors. While generic message passing networks treat all neighbors equally, we utilize a gated \ac{gnn} layer that accounts for per-link channel states, allowing the model to prioritize stronger or weaker links per channel adaptively. 

\smallskip
{\bf Features:} The $l$th gated \ac{gnn} layer (implementing the $l$th message exchange round), applied at the $i$th node, processes its local node features, denoted $\myVec{x}_i^{(l)}$, and outgoing edge features, denoted $\{\myVec{e}_{i\to j}^{(l)}\}_{j\in\mathcal{N}(i)}$. It uses these features to generate messages to all its neighbors, denoted $\{\myVec{m}_{i\to j}^{(l)} \in \mathbb{R}^B\}_{j\in\mathcal{N}(i)}$, and generate updated $B\times 1$ node and edge features, respectively denoted $\myVec{x}_i^{(l+1)}$ and $\{\myVec{e}_{i\to j}^{(l+1)}\}_{j\in\mathcal{N}(i)}$.



{\bf Architecture:}
The layer is comprised of:
\begin{enumerate}
    \item An  {\em encoder} with parameters $\myVec{\theta}_{\rm e}^{(l)}$ that maps  $\myVec{x}_i^{(l)}$ and  $\{\myVec{e}_{i\to j}^{(l)}\}$ into $(i)$ updated edge features $\{\myVec{e}_{i\to j}^{(l+1)}\}$ using layer normalization and an edge-gated \ac{fc} layer with a learnable sigmoid gate; and $(ii)$ a set of messages $\{\myVec{m}_{i\to j}^{(l)}\}_{j \in \mySet{N}(i)}$ using subsequent \ac{film}~\cite{perez2018film}.
    \item An {\em aggregator} with parameters  $\myVec{\theta}_{\rm a}^{(l)}$ that combines the received messages $\{\myVec{m}_{j\to i}^{(l)}\}_{j \in \mySet{N}(i)}$ and applies an \ac{fc} layer to obtain residual update of the node features into $\myVec{x}_i^{(l+1)}$,
\end{enumerate}
The encoder  $\myVec{\theta}_{\rm e}^{(l)}$ and  aggregator   $\myVec{\theta}_{\rm a}^{(l)}$  are shared by all nodes.

\subsubsection{Overall Algorithm}
\begin{figure*}[!t]
  \centering
  \includegraphics[width=0.8\textwidth,keepaspectratio]{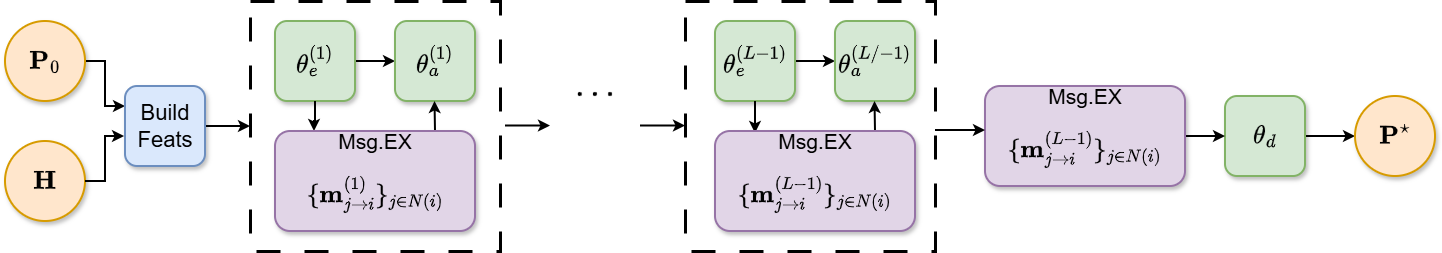}
  \caption{MANET-GNN architecture block diagram.}
  \label{fig:model_architecture}
  \vspace{-0.3cm}
\end{figure*}

The overall algorithm consists of stacking $L-1$ gated \ac{gnn} layers with dedicated input pre-processing. The output of these layers is fed to  a shared decoder that produces the final per-link per-band power allocations $\{p_{i \to j}^{(b)}\}$ using one last message exchange round. The full architecture is illustrated in Fig.~\ref{fig:model_architecture}.

{\bf Input Processing:} 
While the node and edge features generated by each layer are $B\times 1$ vectors, the initial values have different dimensions to embed the required information. Specifically, each initial edge feature vector $\myVec{e}_{i\to j}^{(1)}$ is obtained by stacking the real and imaginary parts of the estimated link \ac{csi}, $\{\hat{\myVec{h}}_{i\to j}^{(b)}\}_{b=1}^{B}$. The input node-features are $B+3\times 1$ vectors, whose  first $B$ entries are set to an initial power division, while the remaining $3$ entries form a one-hot encoding of the role of node $i$ (Tx, Rx, or relay). 

{\bf Output Processing:} 
Every node performs $L-1$ rounds of decentralized message passing with its neighbors, thus updating its local embedding. After these rounds, each node broadcasts its final embedding once to its direct neighbors. 
Then for each $(i \!\to\! j)$, the node $i$ applies a  decoder, comprised of an \ac{fc} layer with $B$ outputs and softplus activation, whose parameters are denoted by $\myVec{\theta}_{\rm d}$, to the concatenation of the final edge embedding and the endpoint embeddings, i.e., to $[\myVec{e}_{i\to j}^{(L)} \,\|\, \myVec{x}_i^{(L)} \,\|\, \myVec{x}_j^{(L)} ]$.  
The resulting non-negative $B\times 1$ vectors are normalized over all edges to produce the local power allocation $[\myMat{P}]_{:,i,:}$.

\subsection{Training}
\label{sec:training}
Our training procedure of MANET-GNN tunes the weights $\myVec{\theta} = [\myVec{\theta}_{\rm d}, \{\myVec{\theta}_{\rm a}^{(l)}, \myVec{\theta}_{\rm e}^{(l)}\}_{l=1}^{L-1}]$, based on the unlabeled data \eqref{eqn:DataSet} and the optimization objective in \eqref{eqn: best power allocation}. To enable unsupervised learning using standard gradient-based training methods, we utilize a relaxation of the objective \eqref{eqn: best power allocation}, and formulate an optimization-oriented loss measure, combined with a noisy-\ac{csi}-aware training scheme.

{\bf Relaxed Objective:} The objective in \eqref{eqn: best power allocation} includes a $\min$ operator over all links which limits accounting for the entire \ac{manet} in training. To mitigate this effect on gradient-based training, we utilize a {\em surrogate rate} objective, defined as 
\begin{equation}
   \tilde{R}(\myMat{P}; \{h_{i\to j}^{(b)}\}) = \sum_{b=1}^{B} 
      \max_{\phi \in \Phi_b} \, 
      {\rm smin}_{(i\to j) \in \phi} R_{i\to j}^{(b)},
      \label{eqn:surrogate}
\end{equation}
where ${\rm smin}$ is the {\em smooth-min} operator, defined as
\begin{equation}
\mathrm{smin}_\tau\bigl(\{s_k\}\bigr)
= -\tau \log \sum_k \exp \Bigl(-\tfrac{s_k}{\tau}\Bigr),
\qquad \tau>0,
\end{equation}
which converges to the true minimum as $\tau\to 0^+$.

{\bf Loss:} 
Our loss  balances two aspects arising from viewing MANET-GNN as a learned optimization solver. The first term evaluates it based on the surrogate rate at its output, i.e., 
\begin{equation}
    \mySet{L}_{\mySet{D}}^{\rm rate}(\myVec{\theta}) = \frac{-1}{|\mySet{D}|} \sum_{d=1}^{|\mySet{D}|} \tilde{R}(\myMat{P}^{(L-1)}_d(\myVec{\theta}) ; \{h_{i\to j,d}^{(b)}\}),
    \label{eqn:UsupLoss}
\end{equation}
where $\myMat{P}^{(l)}_d(\myVec{\theta})$ is the power allocation obtained by applying the decoder $\myVec{\theta}_{\rm d}$ to the output of the $l$th gated GNN layer in MANET-GNN applied to the $d$th sample. 

The second loss term treats MANET-GNN as a decentralized descent method optimizer, in which each message passing round should monotonically increase the resulting rate. This form of regularization is known to facilitate and stabilize learning of \ac{dnn}-aided optimizers \cite{shlezinger2023model}. The resulting loss term is given by 
\begin{align}
  &  \mySet{L}_{\mySet{D}}^{\rm mono}(\myVec{\theta}) = \frac{-1}{|\mySet{D}|(L-2)} \sum_{d=1}^{|\mySet{D}|}  \sum_{l=1}^{L-2} {\rm ReLU}\bigg(\delta -\notag \\
    &\quad \Big(\tilde{R}(\myMat{P}^{(l+1)}_d(\myVec{\theta}) ; \{h_{i\to j,d}^{(b)}\})   \!- \!\tilde{R}(\myMat{P}^{(l)}_d(\myVec{\theta}) ; \{h_{i\to j,d}^{(b)}\})\Big)\bigg) ,
    \label{eqn:MonoLoss}
\end{align}
where $\delta$ is a hyperparameter that enforces a positive margin.

{\bf Noisy-\ac{csi}-Aware Training:} 
The formulation of the loss terms in \eqref{eqn:UsupLoss}-\eqref{eqn:MonoLoss} combined with the surrogate objective in \eqref{eqn:surrogate} enables training MANET-GNN using standard gradient-based learning, with an overall loss of the form
\begin{equation}
     \mySet{L}_{\mySet{D}}(\myVec{\theta}) =  \mySet{L}_{\mySet{D}}^{\rm rate}(\myVec{\theta}) +\lambda  \mySet{L}_{\mySet{D}}^{\rm mono}(\myVec{\theta}),
     \label{eqn:Loss}
\end{equation}
where $\lambda$ is a regularization coefficient. 
Following \cite{alter2025rapid}, we enhance robustness and the ability to cope with noisy \ac{csi} by providing the model noisy \ac{csi} during training (while computing the loss terms using the true \ac{csi}), as a form of adversarial learned optimization~\cite{sofer2025unveiling}. The overall training procedure based on \acl{sgd} is summarized as Algorithm~\ref{alg:train}.

\begin{algorithm}
    \caption{Training MANET-GNN}
    \label{alg:train}
    \SetKwInOut{Initialization}{Init}
    \Initialization{Initial parameters $\myVec{\theta}$; 
    \# batches $Q$; noise $\sigma$; \\
    Learning rate $\eta$; hyperparameters $\delta, L, \lambda$}
    \SetKwInOut{Input}{Input} 
    \Input{Training set  $\mathcal{D}$}   
    {
        \For{${\rm epoch} = 0, 1, \ldots, {\rm epoch}_{\max}-1$}{%
                    Randomly divide  $\mathcal{D}$ into $Q$ batches $\{\mathcal{D}_q\}_{q=1}^Q$;
                    
                    \For{$q = 1, \ldots, Q$}{

                    Apply \ac{gnn} $\myVec{\theta}$ to $\{h_{i\to j,d}^{(b)}\}_{d\in \mathcal{D}_q}$ + $\mySet{CN}(0,\sigma^2)$\;

                    Compute  loss $\mySet{L}_{\mySet{D}_q}$ via \eqref{eqn:Loss};
                    
                    Update  $\myVec{\theta}\leftarrow \myVec{\theta} - \eta\nabla_{\myVec{\theta}}\mathcal{L}_{\mySet{D}_q}(\myVec{\theta})$; \label{stp:update1}
                    }
                    
                    }
        \KwRet{$\myVec{\theta} = [\myVec{\theta}_{\rm d}, \{\myVec{\theta}_{\rm e}^{(l)}, \myVec{\theta}_{\rm a}^{(l)}\}_{l=1}^{L-1}]$}
  }
\end{algorithm}

\subsection{Discussion}
\label{ssec:discussion}

The proposed MANET-GNN framework provides a principled mechanism for meeting the requirements \ref{itm:Decent}-\ref{itm:Noisy}. First, its decentralized message-passing design inherently ensures that each node sets its local power allocations based solely on its local \ac{csi}, thus satisfying \ref{itm:Decent}. The explicit limitation on the number of message-passing rounds guarantees bounded latency and communication overhead, directly addressing \ref{itm:Messages}. The use of \acp{gnn} further enables generalization across unseen network topologies (\ref{itm:transfer}) while the noisy \ac{csi}-aware training in Algorithm~\ref{alg:train} equips the model with robustness to estimated \ac{csi} errors (\ref{itm:Noisy}). 

Several follow-up directions naturally arise from this work. One can potentially expand the optimization task beyond power allocation to also include channel assignment among users, thereby adapting the design to distributed OFDMA systems. Another interesting aspect concerns the optimization objective in \eqref{eqn: best power allocation}, which assumes the common setting where different messages are encoded across different channels. Extending the framework to scenarios where relays can re-encode or combine messages across channels would require reformulating the objective and adapting the architecture accordingly, which we leave for future research.

\section{Experimental Study}
\label{sec:experiments}

\subsection{Experimental Setup}
We evaluate MANET-GNN in \acp{manet} with $|\mySet{V}|=10$ nodes, with $B=6$ channels per-link representing \ac{ofdm} signaling with $B$ subbands\footnote{The source code used in our empirical study, along with the hyperparameters is available at \url{https://github.com/AlterTomer/Decentralized-MANET}}. 
The topologies are randomized as Erdos–Renyi graphs with edge probability $0.5$, leading to relatively dense graphs.   
The multi-subband links were generated using the Rayleigh fading  model. The training set \eqref{eqn:DataSet} contains $4{,}000$ randomly generated topologies with randomized transmitter/receiver placements.

\begin{figure}[t]
\centering
\begin{subfigure}[b]{0.43\textwidth}
  \centering
  \includegraphics[width=\textwidth]{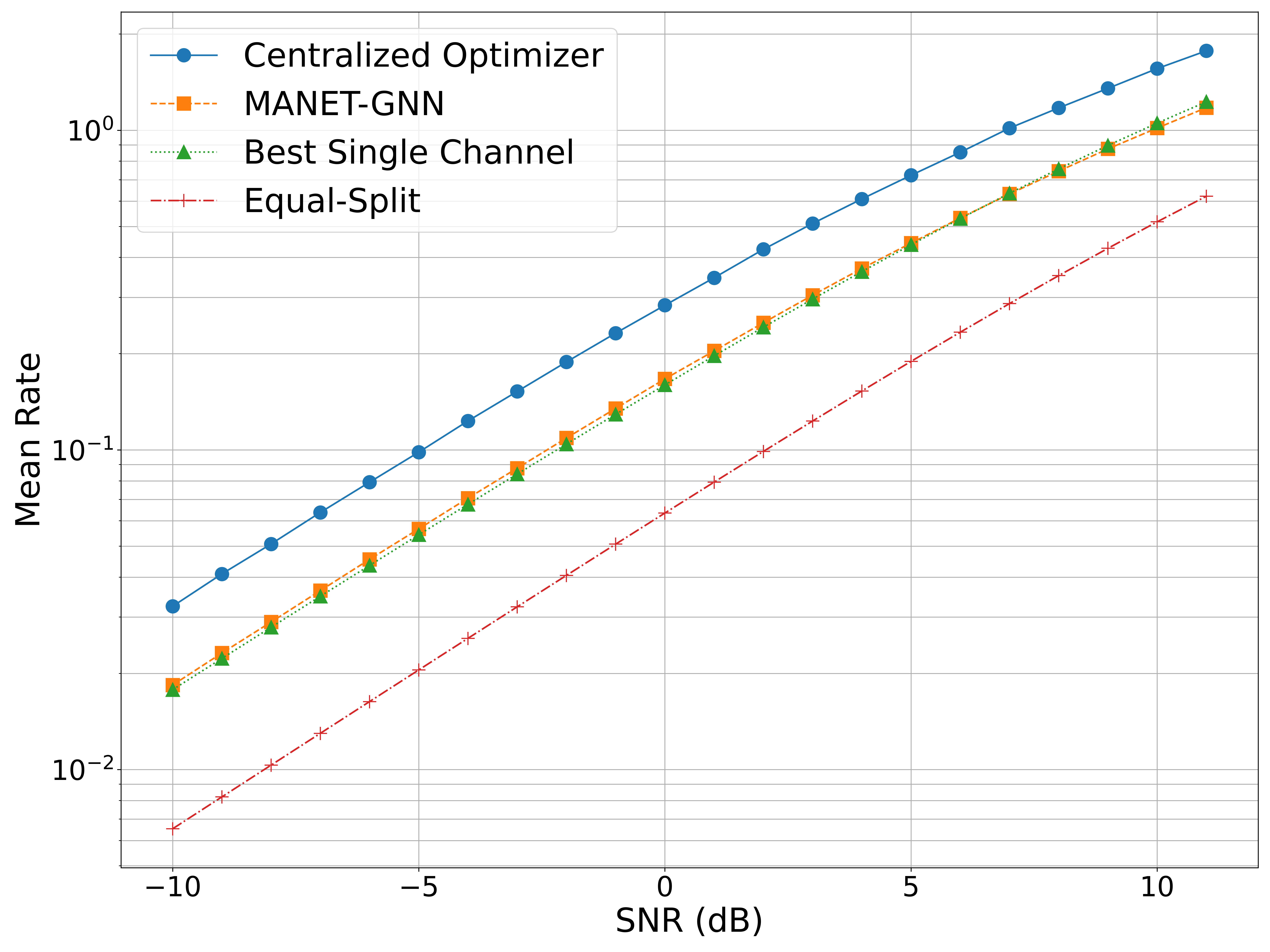}
  \caption{Rayleigh fading, full \ac{csi}.}
  \label{fig:n10-benchmark}
\end{subfigure}\hspace{0.02\textwidth}
\begin{subfigure}[b]{0.43\textwidth}
  \centering
  \includegraphics[width=\textwidth]{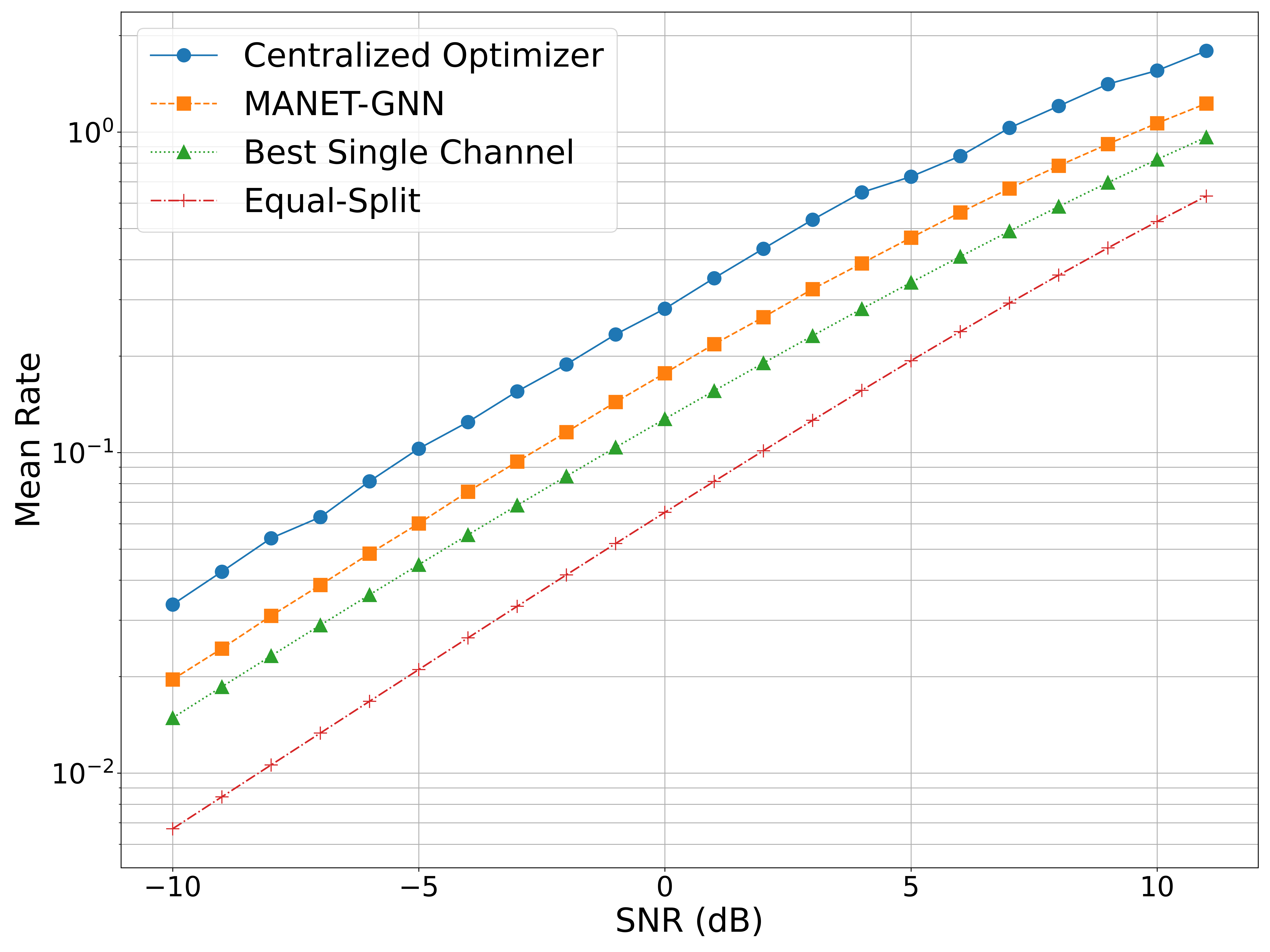}
  \caption{Rayleigh fading, noisy \ac{csi}.}
  \label{fig:estimated_rayleigh_benchmark}
\end{subfigure}
\caption{Mean rate versus \acs{snr} for considered algorithms.}
\end{figure}

We compare the following algorithms:
\begin{enumerate}[label={B\arabic*}]
    \item \label{itm::GNN}\textbf{MANET-GNN} with $L=4$ message exchange rounds.
    \item \label{itm:Opt}\textbf{Centralized optimizer} which tackles \eqref{eqn: best power allocation} using a centralized gradient-based solver (AdamW~\cite{zhou2024towards}) with global \ac{csi} that jointly updates all power variables.
    \item \label{itm:Sing}\textbf{Best single channel}, that assigns all power to a single band $b\in \{1,\ldots,B\}$, selecting the one that optimizes the rate.
    \item \label{itm:Equal} \textbf{Equal-split}, where each node splits its power equally over \emph{all existing} outgoing links and all $B$ bands.
\end{enumerate}
We note that \ref{itm:Opt} is a fully centralized method, while \ref{itm:Equal} is a decentralized local method that involves no collaboration or message exchanges. In principle, \ref{itm:Sing} can be computed in a decentralized manner using the Bellman-Ford algorithm~\cite{bellman1958routing}. However, this requires a number of $B\times 1$ messages (the transmission occurs on all available bands) per node that grows asymptotically at least as $\mathcal{O}(|\mySet{V}|)$  (as opposed to exactly $L$ messages of size $B$ per node as in MANET-GNN), and thus we evaluate it centrally. 



\subsection{Results}
{\bf Full \ac{csi}:} 
Fig.~\ref{fig:n10-benchmark} reports the average achievable rate as a function of the \ac{snr} under full \ac{csi} availability. The proposed \ac{manet}-\ac{gnn} consistently outperforms \ref{itm:Sing} in the low- and medium-\ac{snr} regimes, and achieves comparable performance in the high-\ac{snr} regime. It achieves approximately $85\%$ of the rate achieved by the centralized solver \ref{itm:Opt}, while requiring only a fixed number $L$ of local message-passing rounds and no global \ac{csi}. The naive equal-split baseline \ref{itm:Equal} yields the lowest rates across all regimes, highlighting the necessity of topology and channel-aware coordination.

{\bf Noisy \ac{csi}:}
We next evaluate  the ability of MANET-GNN to cope with noisy \ac{csi} estimates \ref{itm:Noisy}. To that aim, we utilize its \emph{noisy \ac{csi}-aware} training, in which the forward pass uses noisy \ac{csi}, obtained via linear minimum mean-squared error channel estimation using $4$ pilots  for each subband, while the loss is calculated with the corresponding true \ac{csi}. The resulting rate versus \ac{snr} is reported in Fig.~\ref{fig:estimated_rayleigh_benchmark}. There,  we observe that our MANET-GNN decentralized optimizer handles well estimated \ac{csi} as it achieves \(\sim\!80-85\%\) of the achievable rate compared to the centralized optimizer and \(\sim\!35-40\%\) improvement from the best single channel search, when operating with noisy \ac{csi}. 

In Fig.~\ref{fig:estimated_rayleigh_benchmark} the algorithms are provided with noisy \ac{csi}, which MANET-GNN learns to cope with well in its training procedure in Algorithm~\ref{alg:train}. To show that its noisy \ac{csi}-aware training does not notably degrade its performance when the trained MANET-GNN is provided with accurate \ac{csi}, we report in Table~\ref{tab:true_vs_est_csi} the mean rates achieved when MANET-GNN is tested with true full \ac{csi} while trained  with either noisy (estimated) \ac{csi} or with full (true) \ac{csi}. We note in Table~\ref{tab:true_vs_est_csi} that our noisy \ac{csi}-aware training, which was shown in Fig.~\ref{fig:estimated_rayleigh_benchmark} to notably enhance decentralized optimization when provided with estimated \ac{csi}, hardly degrades performance when provided with true \ac{csi}, as  the estimation-aware model achieves \(\sim\!98\%\) of the achievable rate of a model trained using true \ac{csi} only.


\begin{table}[b!]
\centering
\setlength{\tabcolsep}{6pt}
\small
\begin{tabular}{c|ccccc}
\toprule
\textbf{SNR (dB)} & \textbf{-10} & \textbf{-5} & \textbf{0} & \textbf{5} & \textbf{10} \\
\midrule
True CSI      & 0.0200 & 0.0580  & 0.162 & 0.385 & 0.742 \\
Estimated CSI & 0.0196 & 0.05684 & 0.158 & 0.377 & 0.727 \\
\bottomrule
\end{tabular}
\caption{MANET-GNN rate when tested with true \ac{csi} and trained with  true and estimated CSI.}
\label{tab:true_vs_est_csi}
\end{table}

{\bf Scalability:}
We conclude by assessing the ability of MANET-GNN to hold \ref{itm:transfer}, i.e., that it can be applied on different topologies. As the results reported so far utilized multiple different topologies but with the same number of nodes, here we show that MANET-GNN can be trained and tested on topologies that do not differ solely in their connectivity, but also in their sizes. 

\begin{figure}
  \centering
  \includegraphics[width=0.43\textwidth]{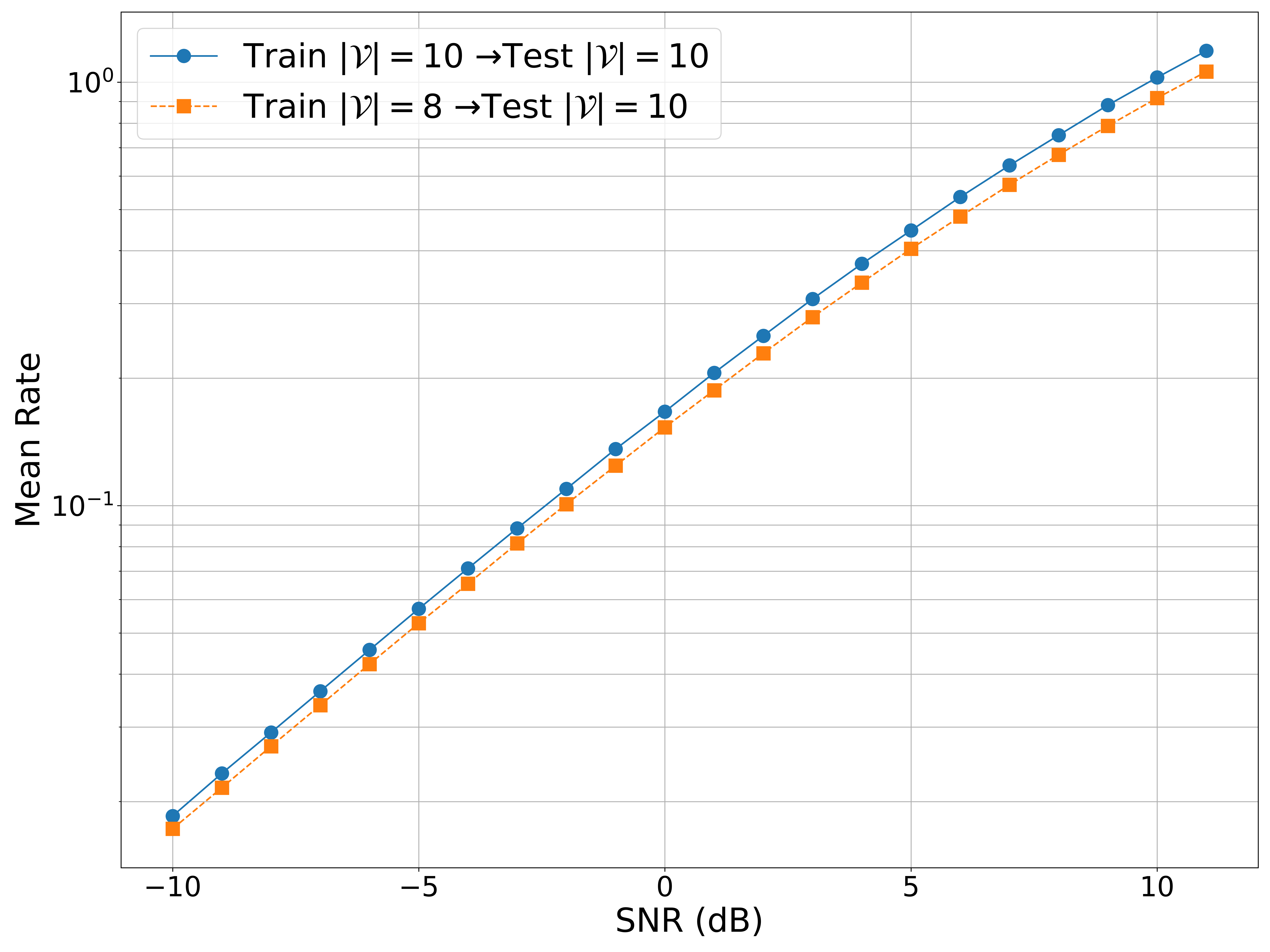}
  \caption{Scalability results: MANET-GNN trained on $|\mySet{V}|\in \{8,10\}$ and tested on $|\mySet{V}|=10$.}
  \label{fig:big_10 small_8}
\end{figure}

To that aim, we train two MANET-GNN models with identical architecture and hyperparameters, both with $B{=}6$: one on \acp{manet} with $|\mySet{V}|{=}8$ nodes and one on $|\mySet{V}|{=}10$. We then evaluate both models on the same test set composed of random \acp{manet} of size $|\mySet{V}|=10$. We compute the mean end-to-end rate using the power allocations predicted by $(i)$ MANET-GNN trained on $|\mySet{V}|{=}10$ (``$10{\to}10$''); and $(ii)$ MANET-GNN trained on $|\mySet{V}|{=}8$ (``$8{\to}10$'').
As shown in Fig.~\ref{fig:big_10 small_8}, the $8{\to}10$ curve closely tracks the $10{\to}10$ curve across the entire \ac{snr} range, indicating strong size-generalization. Any gap that appears is modest, consistent with the increased interference coupling on larger graphs. Overall, these results validate that a MANET-GNN trained on smaller topologies can be deployed on larger ones without retraining as long as the feature dimension is unchanged.

\section{Conclusions}
\label{sec:Conclusions}
In this work, we proposed MANET-GNN, a decentralized \ac{gnn} architecture for power allocation in multi-channel \acp{manet}. By formulating the task as a constrained optimization problem and embedding message-passing operations into the \ac{gnn} design, our approach enables low-latency distributed operation under noisy \ac{csi} while scaling across network sizes and topologies. Numerical evaluations demonstrate that MANET-GNN achieves near-centralized performance, outperforming conventional decentralized baselines.

\bibliographystyle{IEEEtran}
\FloatBarrier
\bibliography{IEEEabrv,refs}
\end{document}